\documentclass[%
reprint,                     
amsmath,amssymb,
aps,
]{revtex4-2}

\usepackage{graphicx}
\usepackage{dcolumn}
\usepackage{bm}

\usepackage{times}
\usepackage{hyperref}

\usepackage{comment}
\begin{document}
	
	\preprint{APS/123-QED}
	
	\title{Entanglement, Squeezing and non-Locality in Filtered  Two-Mode Squeezed Mixed States		
	}
	
	\author{Souvik Agasti}
	\email{souvik.agasti@uhasselt.be}
	
	\affiliation{
		IMOMEC division, IMEC, Wetenschapspark 1, B-3590 Diepenbeek, Belgium
	}%
	\affiliation{
		Institute for Materials Research (IMO), Hasselt University,	Wetenschapspark 1, B-3590 Diepenbeek, Belgium
	}%

	
	\begin{abstract}

	We investigate the entanglement and non-locality between specific spectral components of continuous variable two-mode squeezed mixed states, identifying their limits. 
	These spectral components are selected from output modes using filters commonly employed in optomechanical systems.
	 Both entanglement and non-locality reach their peak when the filters are identical. However, increasing the degree of input squeezing while applying non-identical filters disrupts both entanglement and non-locality, leading to a bell-shaped pattern. Additionally, we provide precise boundaries for entanglement and non-locality.  
		Furthermore, we also evaluate the squeezing of two-mode hybrid quadrature as a measure of entanglement, thereby demonstrating how it remains analogous to logarithmic negativity
		.
	Combined with the filter, the population of two-mode squeezed thermal light influences the angle of a maximally squeezed hybrid quadrature.	
		

	\end{abstract}
	
	\maketitle
	

	\section{Introduction}\label{Introduction}

	Entanglement is a fundamental aspect of the establishment of quantum mechanics which plays a critical role in understanding the Einstine-Podolsky-Rosen (EPR) measurement paradox \cite{EPR_original, EPR_bell}. It had been the key resource in quantum information processing \cite{entanglement_quantum_cryptography}, quantum computation \cite{quantum_computation}, and quantum teleportation \cite{quantum_teleportation}. It quantifies the quantum correlation between continuous-variable states. Gaussian states have mainly been in use for the generation, manipulation, and application of continuous variable entangled states \cite{quantum_teleportation, Entanglement_CV_Plenio}, and therefore, it remained in the intensive interest of the community.
	
	A profound feature of quantum measurement on hybrid systems is its local realism \cite{EPR_original, EPR_bell}, which has extensively been used in the field of quantum information science. The test of quantum non-locality versus local realism still remains a gray area to explore. In this context, Bell's inequality has been considered as a satisfactory condition that all local hidden variables follow \cite{TMSV_nonlocal_Bell_Wigner, bell_homodyne_1, bell_homodyne_2}. Several types of Bell’s inequalities have been explored so far to test quantum nonlocality \cite{bell_CHSH, bell_CH}. All cases use two-body correlation functions as key elements which have been extracted from the two measurement variables at distant places.
	
	The quantum correlated states are generated by using the correlations between a system and ancilla modes, such as two output ports of a beamsplitter \cite{beamsplitter_entanglement_generation}, signal and idler generated through spontaneous parametric down-conversion process (PDC) \cite{SPDC_entanglement_generation}, atomic and cavity modes in cavity quantum electrodynamics \cite{Asjad_Vitali_filter_2, Vitali_Zippilli_NJP, entanglement_Vitali_optomechanics}. The entanglement ensures that the measurement of an observable of one system modifies the state of the other. In all two-mode entangled states, squeezing is observed as an inevitable phenomenon. In fact, in the case of systems with two-mode squeezing (TMS), the squeezing variance can be used as the entanglement criteria \cite{Duan_Inseparability_entanglement, Vitali_Zippilli_NJP}. The squeezing of a quantum state refers to the reduction of fluctuation of a quadrature below the fluctuation of a coherent/vacuum state, which is known as the standard quantum limit (SQL). In the case of TMS, a combination of quadratures can be squeezed, whereas the indevidual modes may not necessarily be squeezed. 

	The two-mode squeezed vacuum (TMSV) states generated by the PDC process have shown that their spectral components have non-trivial quantum correlations. Even though the correlations between the spectral components of PDC-generated single-mode squeezing have been studied and applied before, especially in optomechanical systems \cite{entanglement_Vitali_optomechanics, Vitali_Zippilli_NJP, Asjad_Vitali_filter_2}; however, the quantum correlation between the spectral components of the modes of two-mode squeezing has not been studied so far, which is, therefore, the main motivation of this article. Experimentally, conditional states have been generated through quantum correlations of multi-mode twin-beam states \cite{SPDC_entanglement_generation, correlated_beam_Haderka}.

	In real experimental situations, TMSV faces decoherence due to the presence of a thermal environment. The decoherence and thermalization dynamics of TMSV were studied before by analyzing relative entanglement entropy \cite{Decoherence_TMSV_Tohya, Two_mode_squeezed_vacuum_common_thermal_reservoir}. The TMSV is observed to lose its non-local behavior through decoherence under thermal environment \cite{TMSV_nonlocal_Bell_Wigner}. Besides, it was realized before that the region of entanglement remains unchanged even though the detection efficiency drops \cite{Entanglement_squeezed_thermal}. Even though the interaction of the thermal environment on the TMSV is well studied, the impact of the filter has never been tested so far which we discuss in this article.

	Optical filters are often used to pick up a preferable range of frequency bands from the outputs of quantum systems. 
	In this article, we considered two different types of optical filters, e.g. step and exponential filters. The step filters have already been used before to study the quantum correlation between the modes of optomechanical systems \cite{entanglement_Vitali_optomechanics, Vitali_Zippilli_NJP}. The exponential filters have been used before to analyze the time-dependent physical spectrum of light \cite{filter_2_JOSA} and also in controlling output entanglement in feedback-controlled optomechanical systems \cite{Asjad_Vitali_filter_2}.
	
	One can witness the entanglement through homo/heterodyne measurement of two-mode hybrid quadrature variance. Two-mode quadrature squeezing has proven before to be an equivalence of logarithmic negativity which is a popular measure of entanglement \cite{Duan_Inseparability_entanglement, Simon, Vitali_Zippilli_NJP}.
	Besides, homodyne detection techniques to perform the quadrature phase amplitude measurement, provide quantum correlation which is a platform for testing Bell inequality efficiently \cite{bell_homodyne_1, bell_homodyne_2}.  The analysis enables highlights the loopholes in optical experiments as the violation of Bell's inequality should be taken into care.

	The article is composed as follows. In Sec. II, we discuss the theory of how the filter manipulates the nature of TMSV. We also discuss the impact of detection efficiency in this section. Afterwards, in Sec III we considered a two-mode squeezed thermal light to investigate the impact of thermal population along with filter on entanglement. For experimental realization, we quantify entanglement through the squeezing of hybrid quadrature, where we also see how the temperature controls the squeezed quadrature.


	\section{FILTER on TMSV}

	\subsection{Input TMSV}

	Entangled TMS is typically generated using non-linear optical crystal $(z)$ through the PDC process. Considering $r$ as the amplitude of the squeezing parameter and fixing
	the arbitrary phase to be $\pi/2$ as a reference in this PDC process, the squeezed Bogliubov modes can be represented as 
	
	
	\begin{align}
		\begin{bmatrix}
			X_I^{out}\\
			Y_I^{out}\\
			X_S^{out}\\
			Y_S^{out}
		\end{bmatrix} &= \begin{bmatrix}
			\cosh r & 0 & \sinh r & 0 \\
			0 & \cosh r & 0 & - \sinh r \\
			\sinh r & 0 & \cosh r & 0\\
			0 & - \sinh r & 0 & \cosh r
		\end{bmatrix} \begin{bmatrix}
			X_I^{z}\\
			Y_I^{z}\\
			X_S^{z}\\
			Y_S^{z}
		\end{bmatrix} 
	\end{align}
	
	where $X_{I,S}^{z}(t) = \frac{1}{\sqrt{2}} \left(a_{I,S}^{z}(t) +{a_{I,S}^{z} }^\dagger(t) \right) $ and $Y_{I,S}^{z}(t) = -i\frac{1}{\sqrt{2}} \left(a_{I,S}^{z}(t) -{a_{I,S}^{z} (t)}^\dagger \right) $ are the amplitude and phase quadratures of a multimode bosonic system acts as input to the non-linear crystal (z) of the bipartite systems idler $(I)$ and signal $(S)$. $a_{I,S}^{z} ({a_{I,S}^{z} }^\dagger)$ are the corresponding annihilation (creation) operators, which are related to their spectral components through the Fourier transform: $\tilde{a}^z_{I,S}(\omega) = \sqrt{\frac{1}{2\pi}} \int_{-\infty}^{\infty} e^{i\omega t} a^z(t)$ and $ { \tilde{a^z}^\dagger }_{I,S}(-\omega) = \sqrt{\frac{1}{2\pi}} \int_{-\infty}^{\infty} e^{i\omega t} {a^z}^\dagger(t)$. Therefore, the frequency spectra of the input quadratures are $ \tilde{X}_{I,S}^{z}(\omega) = \frac{1}{\sqrt{2}} \left( \tilde{a}_{I,S}^{z}(\omega) +{ \tilde{a^z}^\dagger }_{I,S}(-\omega) \right) $ and $ \tilde{Y}_{I,S}^{z}(\omega) = -i\frac{1}{\sqrt{2}} \left( \tilde{a}_{I,S}^{z}(\omega) -{ \tilde{a^z}^\dagger }_{I,S}(-\omega) \right) $.  $X_{I,S}^{out}$ and $Y_{I,S}^{out}$ are the outputs obtained after PDC as TMS state. $a_{I,S}^{out} ({a_{I,S}^{out} }^\dagger)$ are the corresponding annihilation (creation) operators for the outputs of the systems $I$ and $S$ which are related to their frequency spectra through Fourier transform, similar to the input quadratures: $ \tilde{X}_{I,S}^{out}(\omega) = \frac{1}{\sqrt{2}} \left( \tilde{a}_{I,S}^{out}(\omega) +{ \tilde{a^{out}}^\dagger }_{I,S}(-\omega) \right) $ and $ \tilde{Y}_{I,S}^{out}(\omega) = -i\frac{1}{\sqrt{2}} \left( \tilde{a}_{I,S}^{out}(\omega) -{ \tilde{a^{out}}^\dagger }_{I,S}(-\omega) \right) $. 
	Details of the PDC process and generation of TMSV are shown in Fig. \ref{Block_Diagram_TMSV}(A). 
	Accepting optical losses before the light enters into the filter, we obtain the modified input field, on which the filter applies, as
	
	\begin{equation}
		\begin{bmatrix}
			X_{I,S}^{out}\\
			Y_{I,S}^{out}
		\end{bmatrix}_{actual} =
		\sqrt{\eta_{I,S} }\begin{bmatrix}
			X_{I,S}^{out}\\
			Y_{I,S}^{out}
		\end{bmatrix}_{ideal} + 
		\sqrt{1 - \eta_{I,S} } \begin{bmatrix}
			X_{I,S}^{vac}\\
			Y_{I,S}^{vac}
		\end{bmatrix} 
	\end{equation}

	where $X_{I,S}^{vac}, Y_{I,S}^{vac}$ are the vacuum amplitude and phase noise operators correspond to the systems I and S, respectively, and $\eta_{I,S}$ are their corresponding optical quantum efficiencies.

	\subsection{Filtered Output Modes}
	
	From the continuous field, one can define and extract many independent optical modes (also, shown in Fig. \ref{Block_Diagram_TMSV}(A)), by selecting different time intervals. 
	Filters on the input select a time interval to extract independent output optical modes
	
	\begin{equation} \label{filter_out_mode}
		a^{K,L}_{I,S} (t) 
		=  \int_{-\infty }^t \mathrm{d}t'	h_{K,L}(t-t') a^{out}_{I,S} (t')
	\end{equation}
	
	where
	$h_{K,L}(t) $ are the filter function of the $K,L$th output modes of the bipartite systems I and S, respectively. 
	The regular output field (without filter) follows the correlation function: $[a_{k}^{out}(t),{a_{k}^{out} }^\dagger(t')] = \delta(t-t') $ where $k\in \{I,S\}$. 
	Equivalently, a generic set of output modes, defined in a time interval, obeys the commutation relation: $[a_{k}^{i} ,{a_{k}^{j}}^\dagger] = \delta_{ij}$. This requires to follow the orthonormality between modes, verified by
	
	\begin{equation}
		\int_{0}^\infty  {h}_i(t) {h}^*_j(t) \, \mathrm{d}t = \delta_{ij}
	\end{equation}
	
	Transforming the filtered output mode, given in Eq. \eqref{filter_out_mode} in the frequency domain, one obtains the generalized expression for the field as
	
	\begin{equation}
		\tilde{a}_{I,S}^{K,L} (\omega) =  \sqrt{2\pi} \tilde{h}_{K,L}(\omega) \, \tilde{a}_{I,S}^{out} (\omega)  = \frac{1}{\sqrt{2\pi}} \int_{-\infty}^\infty \mathrm{d}t \, {a}_{I,S}^{K,L} (t) \, e^{i \omega t} 
	\end{equation}
	
	where $\tilde{h}_k(\omega)$ are the Fourier transform of the filter function, which also satisfies the orthogonality relation
	
	\begin{equation}
		\int_{-\infty}^\infty  \tilde{h}_i(\omega) \tilde{h}^*_j(\omega) \, \mathrm{d} \omega = \delta_{ij}
	\end{equation}

	Here, we choose two different types of explicit sets of filter functions that follow such orthogonality. The type-I is a step filter function, given by
	
	\begin{equation}\label{step_filter_t}
		h_{K,L}(t) = \frac{ \Theta(t) - \Theta(t-\tau_{I,S}) }{\sqrt{\tau_{I,S} }} e^{-i \Omega_{K,L} t }
	\end{equation}
	
	The functions give a set of independent optical modes, distributed around the central frequencies $(\Omega_{K,L})$.
	$\tau_{I,S}$ are the time duration of corresponding systems $I$ and $S$. Therefore, $\tau_{I,S}^{-1}$ are the spectral width of the filters which are related to the mode frequencies as
	
	\begin{equation}\label{periodicity}
		\Omega_{K}-\Omega_{K \pm n} = \pm n\frac{2 \pi}{\tau_{I} }  \hspace{3 mm} \text{and} \hspace{3 mm} \Omega_{L}-\Omega_{L \pm n} = \pm n\frac{2 \pi}{\tau_{S} },  \hspace{3 mm}  \,n \, \text{integer}
	\end{equation}

	Such filter functions had been in use before in \cite{entanglement_Vitali_optomechanics, Vitali_Zippilli_NJP} for filtering out the output modes of optomechanical systems. 

	The Fourier transform of the filter function in Eq. \eqref{step_filter_t}, can be expressed for different frequency intervals as
	
	\begin{equation}\label{step_filter_w}
		\tilde{h}_{K,L}(\omega) =  \sqrt{\frac{\tau_{I,S} }{2 \pi}} e^{i(\omega-\Omega_{K,L})\tau/2} \frac{\sin[(\omega-\Omega_{K,L})\tau_{I,S}/2]}{(\omega -\Omega_{K,L})\tau_{I,S}/2}
	\end{equation}
	
	The central frequencies $(\Omega_{K,L})$ are allowed to vary by an integer multiple of $2\pi/\tau_{I,S}$.


	The type-II consists of an exponential filter function, given by
	
	\begin{equation}\label{exponential_filter_t}
		{h}_{K,L}(t) = \frac{e^{-(1/\tau_{I,S} + i\Omega_{K})t}}{\sqrt{\tau_{I,S}/2}} \Theta(t) ,
	\end{equation}
	
	which has frequency spectral distribution in Fourier space

	\begin{equation}\label{exponential_filter_w}
		\tilde{h}_{K,L}(\omega) =\frac{\sqrt{\tau_{I,S}/\pi}}{1-i \tau_{I,S} (\omega- \Omega_{K,L})} ,
	\end{equation}
	
	The periodicity of this filter function remains the same as Eq. \eqref{periodicity}. Such exponential filters have been used before in \cite{Asjad_Vitali_filter_2, filter_2_JOSA} to analyze the time-dependent spectrum of light in optomechanical systems.

	\subsection{Stationary correlation matrix}
	
	To determine the correlation matrix, we need to determine the quadratures of the filtered output. For both the step and exponential filters, an infinite number of mutually independent output quadratures can also be defined, by tuning their frequencies and bandwidths. However, in the case of two-mode squeezing, the filtered quadratures can be determined as

	\begin{align} \label{quadratures_sqz_freq}
		\begin{bmatrix}
			X_{I,S}^{K,L} (r; t) \\
			Y_{I,S}^{K,L} (-r;t) 
		\end{bmatrix} &= \int_{-\infty}^{t} \mathrm{d}t' {T}_{K,L} (t-t') \begin{bmatrix}
			X_{I,S}^{out} (r;t') \\
			Y_{I,S}^{out} (-r;t')
		\end{bmatrix} 
	\end{align}

	where
	
	\begin{equation} \label{T_mat_freq}
		{T}_{K,L}(t) =\begin{bmatrix}
			\Re({h}_{K,L}) & -\Im({h}_{K,L})\\
			\Im({h}_{K,L}) & \Re({h}_{K,L})
		\end{bmatrix} 
	\end{equation}

	and the dimensionless amplitude and phase quadrature operators of the outputs of the bipartite systems are are defined by $X^{K,L}_{I,S}(r;t) = (a_{I,S}^{K,L}(r;t)+ {a_{I,S}^{K,L}(r;t) }^\dagger)/\sqrt{2}, Y^{K,L}_{I,S}(r;t) = -i (a_{I,S}^{K,L}(r;t)- {a_{I,S}^{K,L} }^\dagger(r;t)  )/\sqrt{2}$. 
	
	One can obtain the correlation matrix of $V(r)$ at a time $t$, with the elements 
	
	\begin{equation}\label{vmat_eliments}
		{V_{ij} (r)} = \frac{1}{2} \langle v_{i}^{out} v_{j}^{out} + v_{j}^{out} v_{i}^{out} \rangle
	\end{equation}
	
	where

	\begin{equation}
		v^{out} = [X_I^{K}(r), Y_I^{K}(r), X_S^{L}(r), Y_S^{L}(r) ]^T
	\end{equation}

	Following Eq. \eqref{quadratures_sqz_freq}, in the frequency domain, the filtered quadratures can be calculated as

	\begin{align} 
		\begin{bmatrix}
			\tilde{X}_{I,S}^{K,L} (r;\omega) \\
			\tilde{Y}_{I,S}^{K,L} (-r;\omega) 
		\end{bmatrix} &= \sqrt{2\pi} \tilde{T}_{K,L} (\omega) \begin{bmatrix}
			\tilde{X}_{I,S}^{out} (r;\omega) \\
			\tilde{Y}_{I,S}^{out} (-r;\omega)
		\end{bmatrix} 
	\end{align}
	
	where $\tilde{T}(t)$ is the Fourier transformed matrix of $	{T}_{K,L}(t)$ given in Eq. \eqref{T_mat_freq}. 
	Likewise Eq. \eqref{vmat_eliments}, by selecting the elements for the squeezing factor $+r$, we estimate the correlation matrix in time domain as $
	V (r;t) =  \begin{bmatrix}
		V_I & V_{corr}^T\\
		V_{corr} & V_S 
	\end{bmatrix}$, 
	where
	
	\begin{subequations} \label{correlation_matrix}
		\begin{align}
			V_I(t) &= \frac{1}{2} \text{Diag}[D_I ,\, D_I ], \\ 
			V_S(t) &= \frac{1}{2} \text{Diag}[D_S ,\, D_S ] \, \text{and} \\
			V_{corr}(t) &= \frac{1}{2}\begin{bmatrix}
				\begin{array}{cccc}
					C_{11} & C_{12} \\
					C_{21} & C_{22} 
				\end{array}
			\end{bmatrix}, 
		\end{align}
	\end{subequations}
	
	Taking into account optical quantum efficiencies $(\eta_{I,S})$, the elements of the steady state correlation matrix for each mode of the filtered output of the bipartite system are
	
	\begin{subequations}
		\begin{align}\label{correlation_matrix_steady_state}
			D_I &= ( 1+2\eta_I\sinh^2 r), \\
			D_S &= (1+2\eta_S\sinh^2 r),  \\
			C_{11} &= -C_{22} = \sqrt{\eta_I \eta_S}  K_{f} \sinh (2 r) ,\\
			C_{12} &= C_{21} = 0
		\end{align}
	\end{subequations}

	where
	
	\begin{subequations}\label{define_K_f}
		\begin{align}
			K_{f} &= \frac{\sin \left[\tau  (\text{$\Omega_K $}-\text{$\Omega _L$})\right]}{ \sqrt{\text{$\tau_I$} \text{$\tau _S$}} (\text{$\Omega_K $}-\text{$\Omega _L$})} 
			\\
			K_{f} &=\frac{2\sqrt{\tau_I \tau _S } (\tau_I+\tau _S )}{\tau_I^2 \tau _S^2 (\Omega_K-\Omega _L)^2+ ( \tau_I +\tau _S)^2}
		\end{align}
		
	\end{subequations}

	$K_{f}$ in Eq. \eqref{define_K_f}(a) stands for the type-I step filters given in Eqs \eqref{step_filter_t} and \eqref{step_filter_w}, where  $\tau = \min[\tau_I,\tau_S]$. 
	Eq. \eqref{define_K_f}(b) gives $K_{f}$ for the filters of type-II, i.e. exponential filters given in Eqs. \eqref{exponential_filter_t} and \eqref{exponential_filter_w}.

	The value of $K_{f}$ drops down from its unit value when the filters are not identical, i.e. $\Omega_K \neq\Omega _L$ and $\tau_I \neq \tau_S$, for both the types of filters.

	\subsection{Two-Mode Entanglement and Squeezing}
	
	The entanglement between two parties can be witnessed by determining the quantity \cite{Simon}
	
	\begin{equation}
		E_N = \max[0, -\ln 2\nu^-],
	\end{equation}
	
	where
	
	\begin{equation}
		\nu^- = \sqrt{\frac{ \Sigma (V) + \sqrt{\Sigma (V)^2 - 4 \det (V)} }{2}}
	\end{equation}
	
	where $ \Sigma(V ) = ( \det V_I + \det V_S - 2 \det V_{corr} )$

\begin{figure}[t!]
		\includegraphics[width= 1 \linewidth]{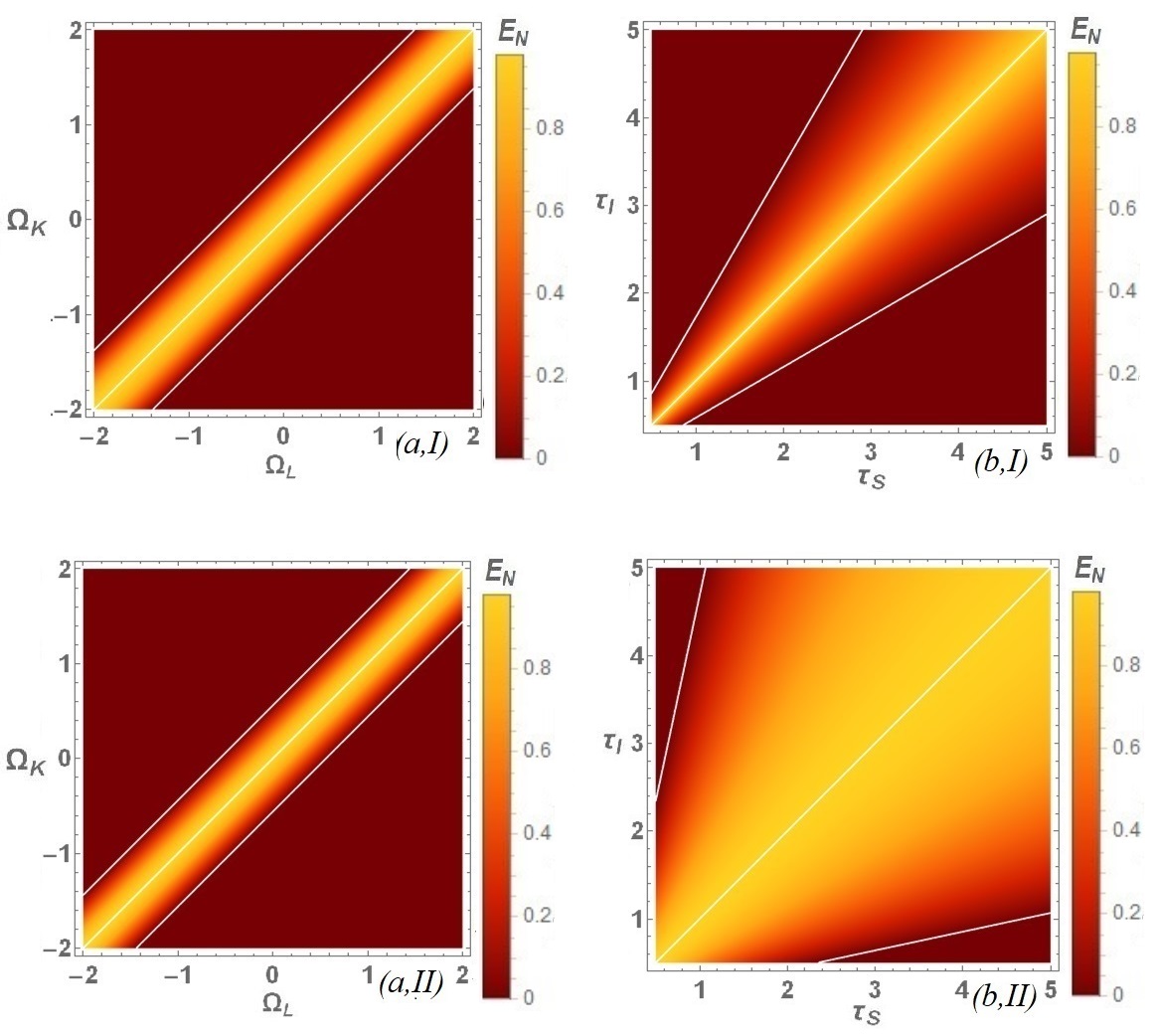}
		\caption{ Entanglement between filtered outputs for the variation of parameters of (a) central frequencies for  $ \tau_I= \tau_S=2 $ a.u.) (b) filter linewidths for  $ \Omega_L = \Omega_K= 1$ a.u.), for (I)step filter and (II) exponential filter. Fixed parameter is $r=1$. }\label{ENvsTauOmega}
	\end{figure}

	We visualize how the entanglement between the filtered outputs of the bipartite system changes with filter parameters in Fig. \ref{ENvsTauOmega}. Firstly, we vary filter frequencies for both the parties, maintaining linewidth unchanged in Fig. \ref{ENvsTauOmega}(a), and observe the entanglement to be maximized when the central frequencies match $(\Omega_K=\Omega_L)$ for both the types of filters. It also shows a drastic drop in the entanglement when the central frequencies mismatch $(\Omega_K\neq\Omega_L)$. Similar to that, the entanglement also shows to be maximized when the bandwidth matches $(\tau_I =\tau_S)$ in Fig. \ref{ENvsTauOmega}(b), and drops down with mismatch. Anticipating that, however, interestingly we see, unlike central frequencies, the region of entanglement becomes wider with the increment of time duration $(\tau_{I,S})$, i.e. the decrement of spectral width. Besides, we observe a difference between type-I and type-II filters, i.e. compared to the type-I filter, the region of entanglement becomes narrower for the variation of central frequencies and wider for the variation of bandwidths for the type-II filter. The impact of filter frequencies and their bandwidths can simply be taken into account with the factor $K_f$ given in Eq. \eqref{define_K_f}.

\begin{figure}[t!]
		\includegraphics[width= 1 \linewidth]{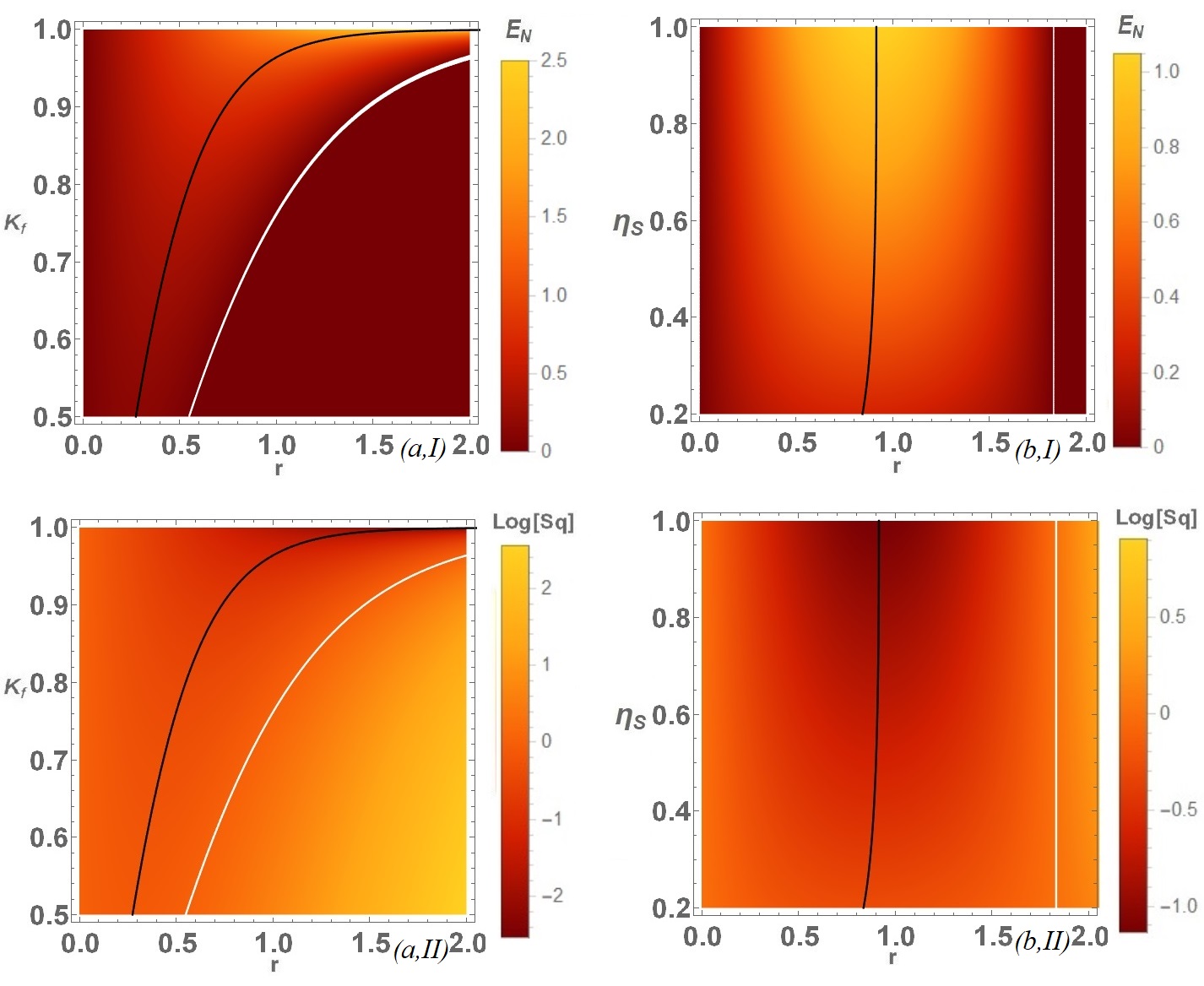}
		\caption{ 
			Entanglement (I) and maximally optimized squeezing (II) between filtered outputs for the variation of parameters: (a) $K_f$ and $r$
			for fixed optical quantum efficiencies $ \eta_I = 0.9, \eta_S = 0.98$, 	
			and (b) $\eta_S$  and $r$ for fixed $K_f = 0.95$ and $ \eta_I = 0.9$. 
			Black lines stand for the maximal entanglement and squeezing, and white lines indecate the boundary of entanglement and SQL. 
		}\label{ENSQTMSV}
	\end{figure}

	The mismatch of frequencies and bandwidths eventually decreases $K_f$ given in Eq. \eqref{define_K_f}. Fig. \ref{ENSQTMSV}(I) shows how the entanglement increases initially and decreases further with the increment of input degree of squeezing $(r)$, making a bell shape. 
	The entanglement shows to have an upper cutoff limit for the squeezing parameter $\left( E_N (r \ge r^{EN}_{ucf}) =0 \right)$, which is determined by $K_f$ as
	
	\begin{equation}
		\tanh(r^{EN}_{ucf})=  K_f.
	\end{equation}
	
	Fig. \ref{ENSQTMSV}(a,I) shows the bell shape becomes narrower and the entanglement is also observed to destroy with the decrement of $K_f$ from its unit value. 
	Therefore, the upper cutoff shrinks with the mismatch of filter parameters. It remains unchanged irrespective of optical quantum efficiencies ($ \eta_{I,S}$), even though the entanglement between the filtered output modes has shown to have a significant dependency on them (Fig. \ref{ENSQTMSV}(b,I)). Earlier, it was realized that the entangled state always remains entangled irrespective of detection efficiency \cite{Entanglement_squeezed_thermal}. The phenomenon remains unchanged even after applying filer on output, as the entanglement is seen to be robust against the decrement of detection quantum efficiency. However, the entanglement was reduced with the reduction of detection efficiency, also shown in Fig. \ref{ENSQTMSV}(b,I). 
	The entanglement reaches its' maximum for the input squeezing  $r^{EN}_{max}$, where

	\begin{widetext}
		\begin{align}
			\tanh (2r^{EN}_{max} ) =  \left(\frac{ \sqrt{4 \eta _I \eta _S K_{f}^2 \left( 2 \eta _I \eta _S + \left(\eta _S+\eta _I\right) \sqrt{  \left(1-K_{f}^2\right) \eta _I\eta _S } -\eta _I \eta _S   K_{f}^2 \right)}}{\left(2 \eta _I \eta _S+  \left(\eta _S+\eta _I\right) \sqrt{  \left(1-K_{f}^2\right) \eta _I\eta _S }\right) }\right).
		\end{align}
	\end{widetext}

	$r^{EN}_{max}$ can be maximized when the condition $\eta_I = \eta_S$ satisfies. In that situation, the entanglement profile becomes uniformly bell-shaped and the maximum is obtained at $r^{EN}_{max} = \frac{1}{2} r^{EN}_{ucf}$. 
	
	
	The filtered arbitary composite quadrature operator (angle $\phi_I$ and $\phi_S$, respectively) of the two-party system for the  output  is 
	
	\begin{equation}\label{composite_quadrature}
		X^{KL(\phi_I\phi_S)}_{IS} = \frac{1}{\sqrt{2}} ( e^{-i\phi_I} a^{K}_I +  e^{i\phi_I} {a^{K}_I}^\dagger +  e^{-i\phi_S} a^{L}_S +  e^{i\phi_S} {a^{L}_S}^\dagger ) .
	\end{equation}
	
	We evaluate the fluctuation of the two-party filtered quadrature operator:
	
	\begin{equation}
		S_q(X^{KL(\phi_I\phi_S)}_{IS}) = \frac{1}{2} \langle \lbrace X^{KL(\phi_I\phi_S)}_{IS}, X^{KL(\phi_I\phi_S)}_{IS} \rbrace \rangle 
	\end{equation}
	
	The maximally squeezed quadrature is realized for $\phi_I+\phi_S = \pi$. The squeezing becomes maximum for the value of $r$:
	
	\begin{equation}
		\tanh(2r_{max}^{SQ}) =  \frac{2 \sqrt{\eta_I \eta_S} K_f}{\eta_I + \eta_S}.
	\end{equation} 
	
	The squeezing of the hybrid quadrature destroys as it goes beyond the standard quantum limit (SQL), when the input squeezing increases to $r_{ucf}^{SQ}= 2r_{max}^{SQ}$ 
	. Note that the cutoff limit of squeezing $(r_{ucf}^{SQ})$ is dependent on optical quantum efficiencies, and therefore different than $r_{ucf}^{EN}$. Both limits match together only when the quantum efficiencies match $(\eta_I=\eta_S)$.

\begin{figure}[t!]
		\includegraphics[width= 1 \linewidth]{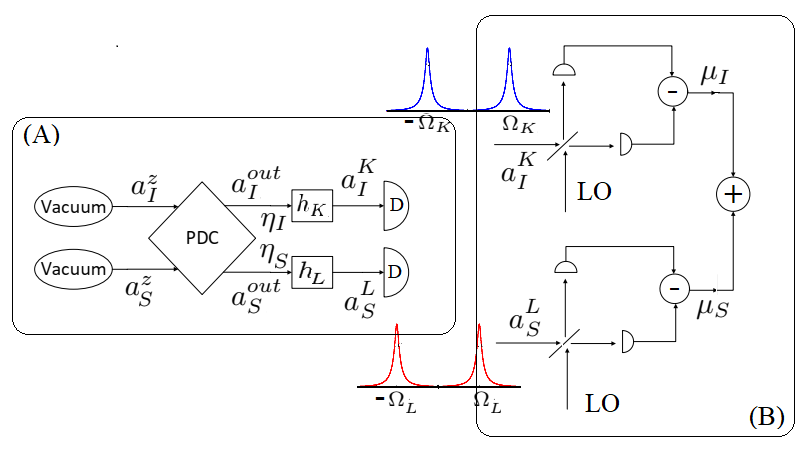}
		\caption{ Block diagram of the detection of filtered TMSV: (A)  TMSV is generated using the parametric down-conversion (PDC) process. Optical filters are applied on two-mode squeezed output before being detected at D. (B) Two-mode homodyne detection and the spectral modes are combined with weightage. LO is the local oscillator. }\label{Block_Diagram_TMSV}
	\end{figure}

	Squeezing and entanglement are two very related concepts, as we know that the squeezing variance can be used as entanglement criteria \cite{Duan_Inseparability_entanglement, Vitali_Zippilli_NJP}. The squeezing variance is estimated for the hybrid quadratures constructed with a weighted mixture of the output photocurrent as shown in Fig. \ref{Block_Diagram_TMSV}(B). The weighted quadrature is
	
	\begin{multline}\label{weighted_quadrature}
		X^{KL(\phi_I\phi_S)}_{IS(\mu_I\mu_S)} = \frac{1}{\sqrt{\mu_I^2+\mu_S^2}} \bigg[\mu_I e^{i\phi_I} a^{K}_I + \mu_I e^{-i\phi_I} {a^{K}_I}^\dagger \\
		+ \mu_S e^{i\phi_S} a^{L}_S + \mu_S e^{-i\phi_S} {a^{L}_S}^\dagger \bigg]
	\end{multline}
	
	where $\mu_I,\mu_S$ are the weight parameters introduced as the scaling of two systems, considered to be real and positive. 
	
	
	Adapting the balance between weights, we obtain maximal squeezing (see Appendix \ref{optimization_squeezing})

	\begin{multline}
		S_q(X^{KL}_{IS})_{optimized} = 1+ (\eta_I+\eta_S) \sinh ^2(r) \\
		-\sinh (r) \sqrt{4 \eta_I \eta_S K_f^2 \cosh ^2(r)+(\eta_I-\eta_S)^2 \sinh ^2(r)}
	\end{multline}
	
	The minimized squeezing follows the same profile of entanglement can be visualized in Fig. \ref{ENSQTMSV}(II). The plots also show the boundary where the squeezing goes beyond SQL which lies the same with the entanglement limits. Therefore, the squeezing is observed to be maximized at $r= r^{EN}_{max}$. 
	One can obtain two modes squeezing beyond SQL till $r\le r^{EN}_{ucf}$. After this limit, the fluctuation can not be suppressed below SQL. At this limit the weight ratio becomes $\frac{\mu_I}{\mu_S}  =  \sqrt{\frac{\eta_S}{\eta_I}}$.

	\begin{figure}[t!]
		\includegraphics[width= 1 \linewidth]{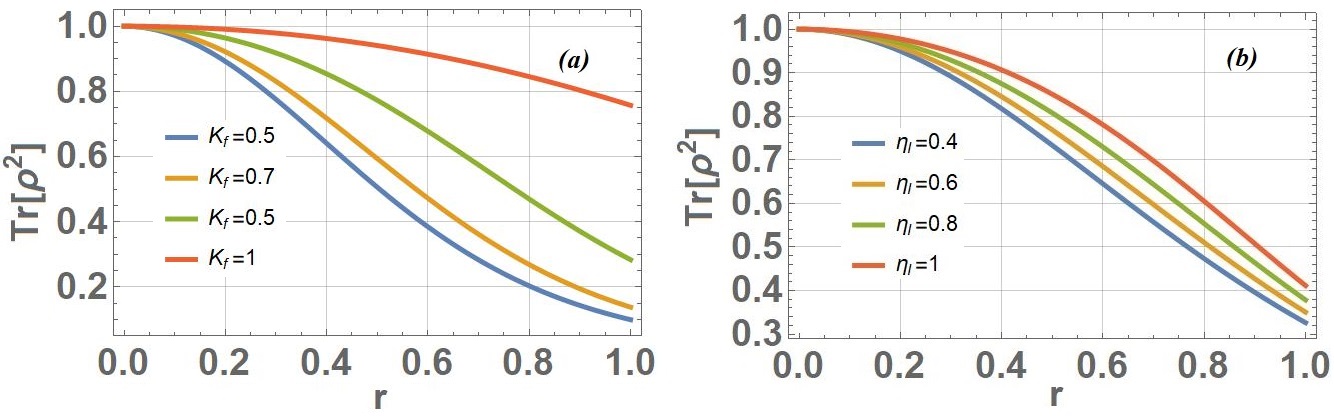 }
		\caption{ Purity of the TMSV as a function of $r$, for (a) fixed optical quantum efficiencies $ \eta_I = 0.9, \eta_S = 0.98$, and (b) fixed $K_f = 0.95$ and $ \eta_I = 0.9$. }\label{purity_TMSV}
	\end{figure}

	\begin{figure}[t!]
		\includegraphics[width= 1 \linewidth]{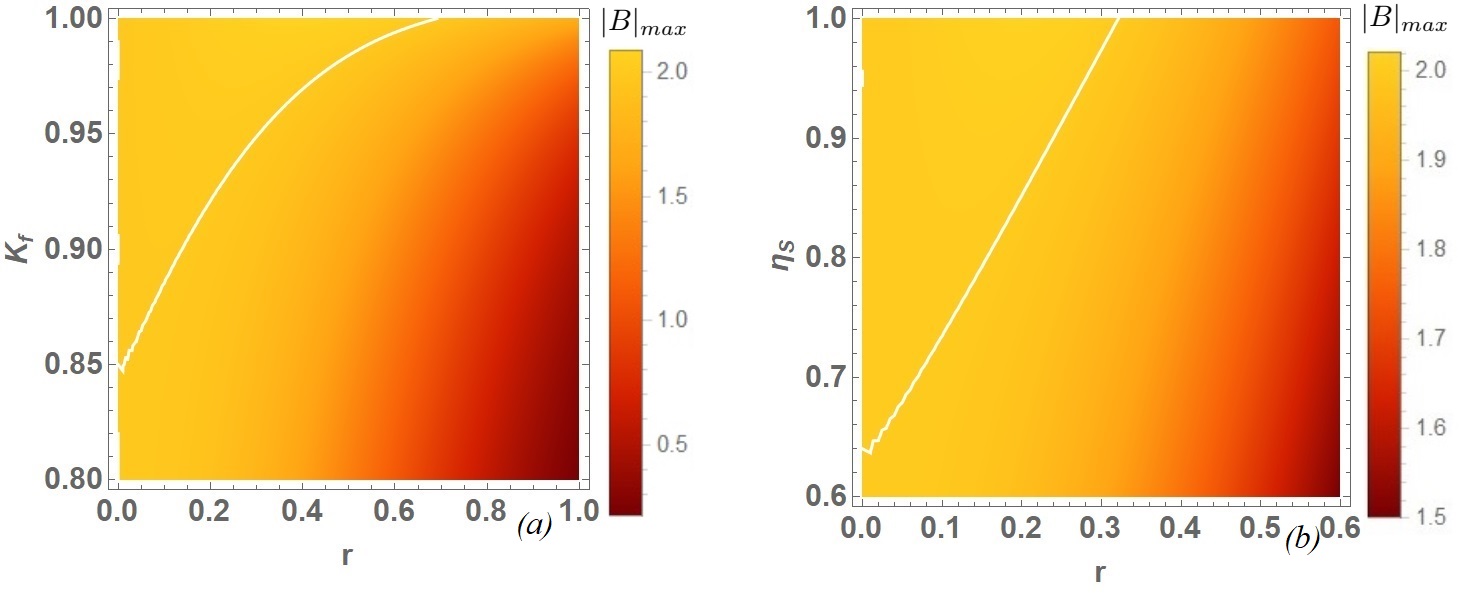}
		\caption{ The maximal value of Bell function as a function of (a) $K_f$ and $r$ for $\eta_I=0.9, \eta_S=0.98$, 		
			and (b)  $\eta_S$ and $r$ by fixing $\eta_I = 0.9, K_f = 0.95$. }\label{Bell_TMSV}
	\end{figure}


	\begin{figure}
		\includegraphics[width= 0.9 \linewidth]{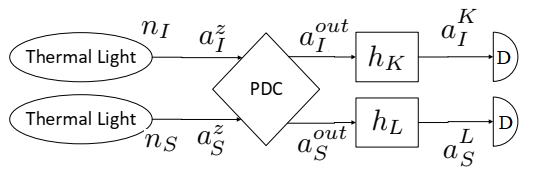}
		\caption{ Block diagram of the generation and detection of filtered TMS thermal light. The only difference with Fig. \ref{Block_Diagram_TMSV} is the input source is replaced by a thermal light. }\label{Block_Diagram_thermal}
	\end{figure}
	
	\begin{figure}
		\includegraphics[width= 1 \linewidth]{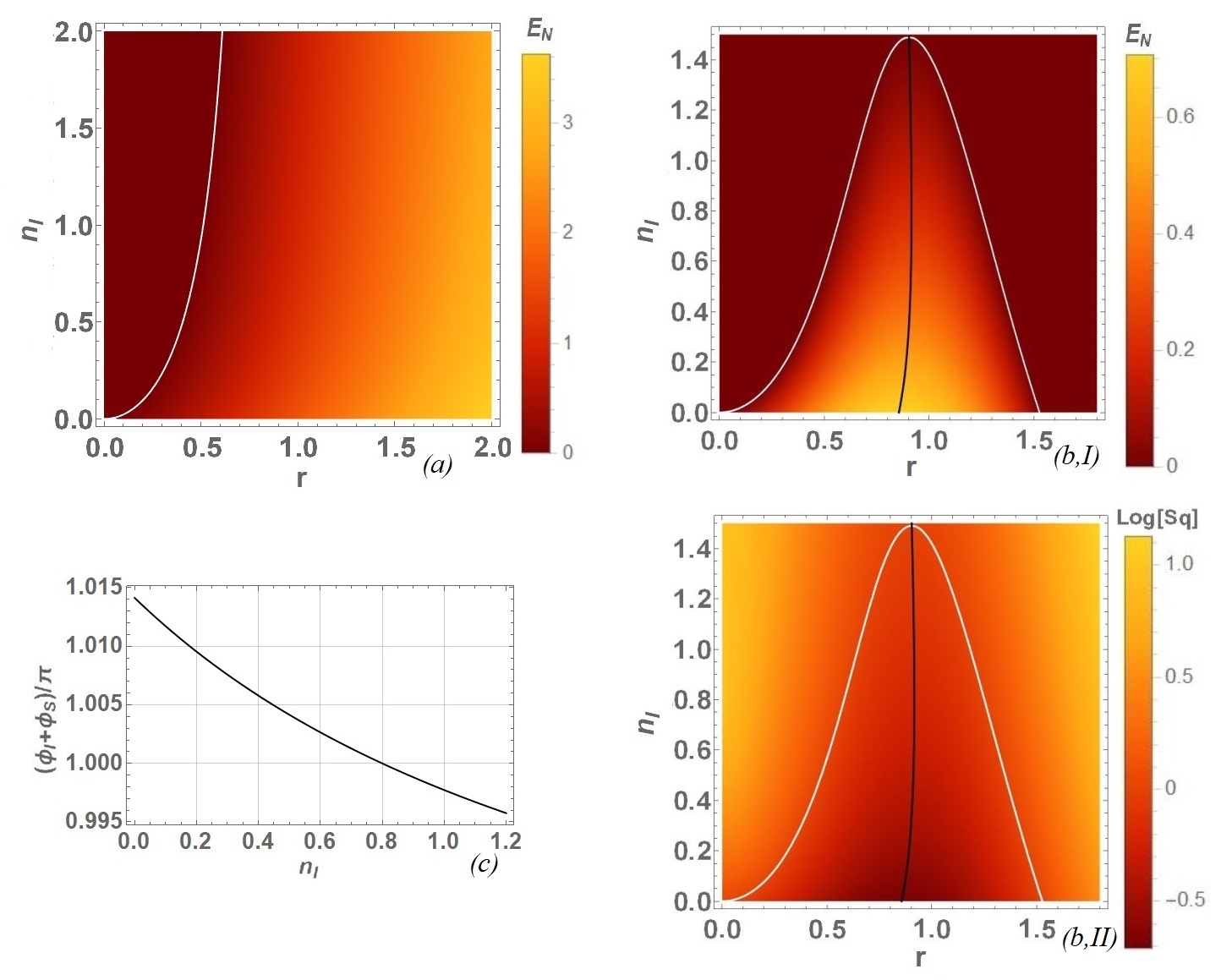}
		\caption{ Entanglement of the filtered outputs for (a) identical filters: $ K_f = 1, L_f = 0 $ which shows the lower entanglement boundary of $(r)$. Entanglement (b,I), maximally optimized squeezing (b,II) and Maximally squeezed hybrid quadrature as a function of thermal population $n_I$ (c)			
			for	non-identical filters $K_f = 0.95, L_f = 0.095 $ which gives both the lower and upper limits of entanglement. 		
			Black lines stand for the maximal entanglement and squeezing, and white lines indecate the boundary of entanglement and SQL. 
		For all the cases $n_S= 0.8$ and the optical quantum efficiencies $(\eta_{I,S})$ are considered to be 1.
	 }\label{EntVSthrmPop_nI0_5tauI2_5wK1_2}
	\end{figure}

	\begin{figure}[t!]
		\includegraphics[width= 0.8 \linewidth]{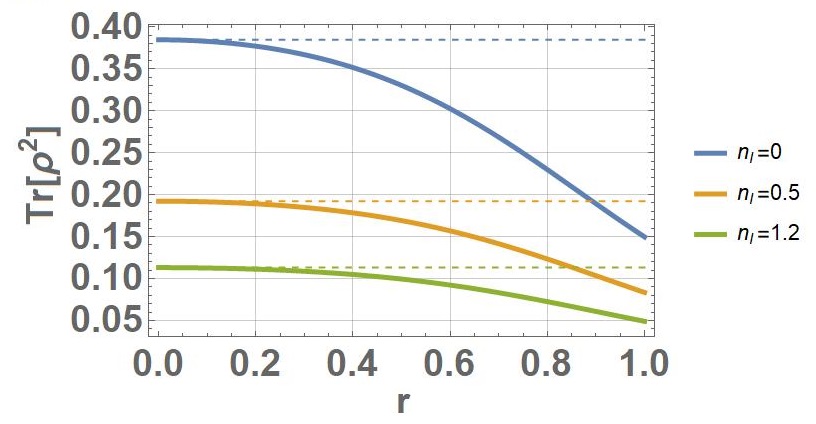 }
		\caption{ Purity of the TMS thermal state as a function of $r$ are represented by thick solid lines for fixed $K_f = 0.95$, $ L_f = 0.095$ and $n_S = 0.8$. Thin dashed lines represent purity of that state without/idetical filters ($K_f = 1, L_f = 0 $). All optical quantum efficiencies are fixed to 1. }\label{purity_thermal}
	\end{figure}

	\begin{figure}
		\includegraphics[width= 1 \linewidth]{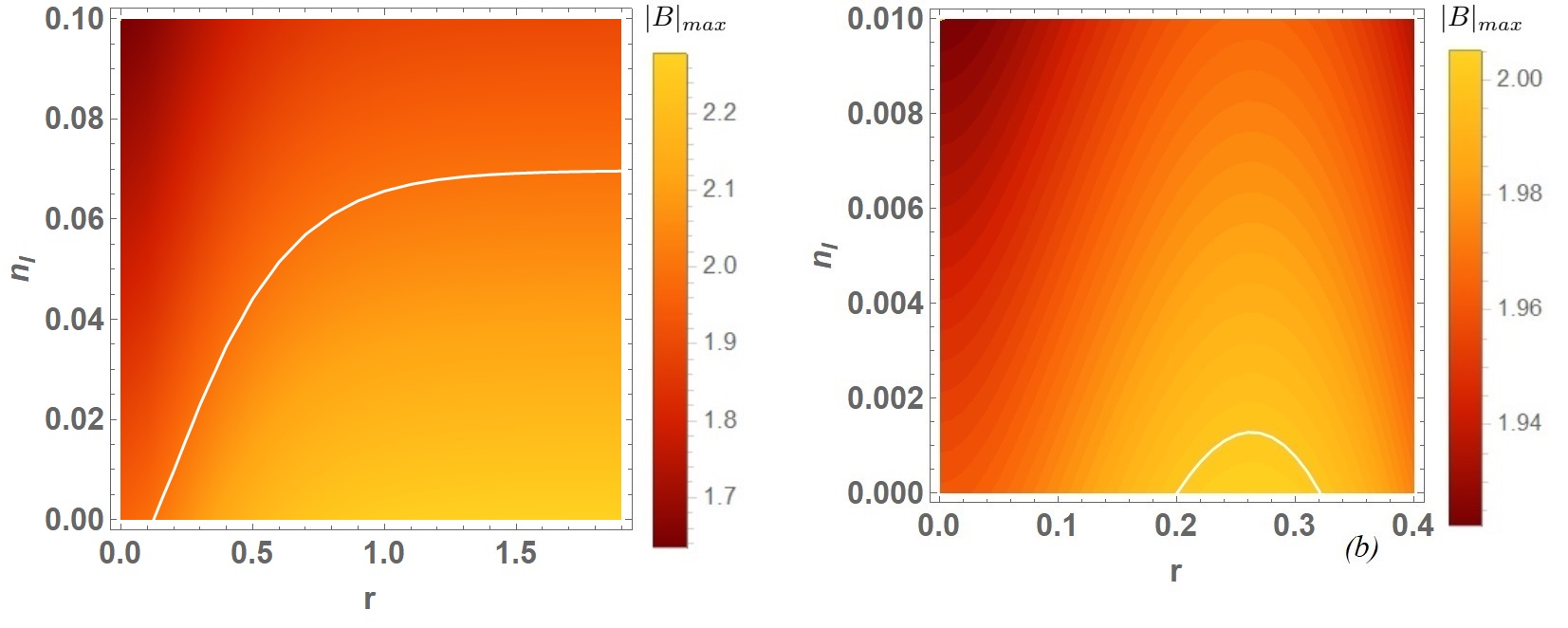}
		\caption{ The maximal value of Bell function as a function of $n_I$ and $r$ for idential filters in  (a), 
			and (b) non-identical filters: $K_f = 0.95$, $ L_f = 0.095$. Both the cases $n_S = 0.01$.
		 The optical quantum efficiencies $(\eta_{I,S})$ are always considered to be 1. }\label{Bell_Thermal}
	\end{figure}

	\subsection{Purity of State}
	
	The entanglement becomes necessery and sufficient criteria of non-locality for pure states. However, in case of mixed states it is not true, and therefore, it is essential to investigate the purity of the states. 
	The purity of states are determined from their density matrix \cite{Purity_Gaussian_states}. 	
	The Gaussian states are characterized by their Wigner function, which is expressed by
	
	\begin{equation}
		W( {u_M}) = \frac{1}{ \pi^2 \sqrt{\det[ {V(r)}]}} \exp \left[ -\frac{1}{2} {u_M}^T {V(r)}^{-1} {u_M}\right].
	\end{equation}

	where  ${u_M} = [Q_I, P_I, Q_S, P_S ]^T $ is the common vector of the fluctuation of two modes. The purity of the state ($\rho$) is measured by 
	
	\begin{equation}
		\mathrm{Tr}[\rho^2] = \frac{\pi^2}{4} \int_R  W^2({u_m}) \mathrm{d} u_m
	\end{equation}
	
	which gives
	
	\begin{equation}
		\mathrm{Tr}[\rho^2] = \frac{1}{ |C_{11}^2 + C_{12}^2 - D_I D_S| } 
	\end{equation}
	
	which is unit valued for vacuum and TMSV, clearly indicating pure states. However, non-identical filters on output destroy the purity, observed in Fig. \ref{purity_TMSV}(a). The mixness of the state is also observed to increase with the decrement of detection quantum efficiency (Fig. \ref{purity_TMSV}(b)). Unlike entanglement, in both cases, the purity is always observed to reduce with the increment of input squeezing (r).

	\subsection{Quantum Non-Locality}
	
	Quantum nonlocality of two-party entangled continuous variables is a natural interest of measurement. A given state is considered to be non-local when it violates Bell's inequality. The Bell's inequality is defined to be $|B|_{max}\leq 2$ using maximum value of the Bell's function $(B)$, which can be expressed in terms of Wigner functions as	
	
	\begin{equation}
		B = \frac{\pi^2}{4} \left[ W(u_M^{00}) +W(u_M^{01}) W(u_M^{10}) - W(u_M^{11})          \right]
	\end{equation}
	
	where ${u^{mn}_M} = [Q^m_I, P^m_I, Q^n_S, P^n_S ]^T $. Such type of Bell's inequality was first described by Clauser, Horne, Shimony and Holt \cite{bell_CHSH}. In quantum mechanics, joint probability distributions are often been expressed using the Q-function or Wigner function. Therefore, the Wigner function has always remained to be useful to test quantum non-locality. 
	Even though violation of all possible Bell's inequalities are not necessary criteria of being nonlocal; in the case of TMS states, violation of Bell's inequality confirms non-locality \cite{Banaszek_TMSV_bell, Banaszek_TMSV_bell_PRL}.

	Quantum mechanically, the field is considered to be nonlocal when $|B|_{max}$ reaches above 2 and the non-locality gets stronger as $|B|_{max}$ becomes larger. 
	We determine $|B|_{max}$ numerically for the non-identical filters in Fig. \ref{Bell_TMSV}. Likewise entanglement or squeezing profile, we observe the Bell function to drop down with the initial degree of squeezing $(r)$. 
	Also, the region of nonlocality reduces rapidly with the reduction of $K_f$ from its unit value (Fig. \ref{Bell_TMSV}(a)). However, unlike entanglement, even in the case of identical filters, the region of non-locality is limited concerning the initial degree of squeezing. Besides, we also realize from Fig. \ref{Bell_TMSV} (b) that the region of non-locality reduces with the reduction of output quantum efficiency, which is a distinct difference compared to the entanglement profile in Fig. \ref{ENSQTMSV}(b).
	Overall, the region of quantum non-locality appears to lie within the territory of entanglement, which
	can be justified by the mixedness of states.

	For pure states 
	the violation of Bell inequalities occurs if and only if the state is entangled. However, for mixed states, entanglement
	is necessary but not sufficient to ensure the violation of Bell inequality \cite{mixed_entanglement_nolocal}. The investigation of purity of the state appears in this context while studying the violation of Bell's inequality to compare with entanglement limits.

	\section{ TWO MODE SQUEEZED THERMAL STATE}
	
	\subsection{Entanglement and Squeezing}
	
	
	We furthermore considered two-mode squeezed thermal light to study the impact of filters on it. 
	The basic block diagram is given in Fig. \ref{Block_Diagram_thermal}. Such states, even though have already been studied before in \cite{Entanglement_squeezed_thermal, Two_mode_squeezed_vacuum_common_thermal_reservoir}, the impact of filter on squeezed output have not been investigated. Therefore, adding up filters given in Eqs. \eqref{step_filter_t}, \eqref{exponential_filter_t} into the model of squeezed thermal states, gives us interesting features to be studied. The effective thermal population of the multimode input light is controlled by the temperature and the frequencies that we are interested in.  
	The elements of the correlation matrix of the Eq. \eqref{correlation_matrix} gives

	\begin{subequations}
		\begin{align}
			D_I &= B +  A \cosh(2r) \\
			D_S &= - B +  A\cosh(2r) \\
			C_{11} &= -	C_{22} = A K_f \sinh (2 r)\\
			C_{12} &= C_{21}=  - B  L_f \sinh (2r)
		\end{align}
	\end{subequations}

	where 
	
	\begin{subequations}
		\begin{align}
			A =& n_I+n_S +1, \,\, B = n_I - n_S ,\\
			L_f =& \frac{ 1- \cos \left(\tau \left(\Omega _K-\Omega _L\right)\right)  }{ \sqrt{\text{$\tau_I$} \text{$\tau _S$}} \left(\Omega _K-\Omega _L\right)} \text{for Type-I step filters: Eq. \eqref{step_filter_t}} \nonumber \\
			=& \frac{2 (\tau_I \tau _S)^{3/2} (\Omega _K-\Omega _L)}{\tau_I^2  \tau _S^2 (\Omega _K-\Omega _L)^2+ ( \tau_I+ \tau _S)^2} \nonumber\\
			&\text{for Type-II exponential filters: Eq. \eqref{exponential_filter_t}}
		\end{align}
	\end{subequations}

	where  $\tau = \min[\tau_I,\tau_S]$ and $\kappa_I$ and $\kappa_S$ are the rates of dissipation, $n_I$, and $n_S$ are the thermal quanta of their corresponding reservoir for the systems $I$ and $S$, respectively. Since the entanglement limits have not been changed due to optical efficiencies, we have chosen the detection efficiency to be fixed and maximized for this section of the article. 
	One can always determine the steady state correlation matrix and find the Eq. \eqref{correlation_matrix} for a situation when the reservoir is a vacuum. 

	In the case of thermally populated reservoirs, the entanglement at steady state shows to have both upper and lower cutoff limits for the input squeezing parameters ($r^{EN}_{ucf}$ and $r^{EN}_{lcf}$, respectively), representing separability condition. Both the upper and lower limits are determined as
	
	\begin{widetext}
		\begin{subequations} \label{upper_lower_limit_thermal}
			\begin{align}
				\cosh(2r_{ucf}^{EN}) &=  \frac{A+\sqrt{\left(A^2 \left(K_f^2-1\right)+B^2 L_f^2\right) \left(A^2 K_f^2+B^2 \left(L_f^2-1\right)+1\right)+A^2}}{A^2 \left(1-K_f^2\right)-B^2 L_f^2} \\
				\cosh(2r_{lcf}^{EN}) &= \frac{A-\sqrt{\left(A^2 \left(K_f^2-1\right)+B^2 L_f^2\right) \left(A^2 K_f^2+B^2 \left(L_f^2-1\right)+1\right)+A^2}}{A^2 \left(1-K_f^2\right)-B^2 L_f^2}
			\end{align}
		\end{subequations}
	\end{widetext}
	
	The steady-state entanglement between two filtered modes has been plotted in Fig. \ref{EntVSthrmPop_nI0_5tauI2_5wK1_2} for (a) identical filters $(\Omega_{K}=\Omega_{L})$ and $(\tau_{I}=\tau_{S})$, and therefore $K_f= 1$ and $L_f= 0$, and for (b,I) non-identical filters: $\Omega_{K}\neq\Omega_{L}$ and $\tau_{I}\neq\tau_{S}$ resulting $K_f < 1$ and $L_f>0$. As witnessed from the Eq. \eqref{upper_lower_limit_thermal}, the upper limit of squeezing factor $(r_{ucf}^{EN})$ blows up for the perfect homodyne detection when the filters are identical. However, one can anticipate a lower limit of the squeezing parameter even in case of identical filters or no filters applied on outputs (Fig. \ref{EntVSthrmPop_nI0_5tauI2_5wK1_2}(a)). The entanglement declines with the increment of thermal populations of the reservoirs. This moreover ensures an extension of the lower cutoff $(r_{lcf}^{EN})$ in both the cases Fig. \ref{EntVSthrmPop_nI0_5tauI2_5wK1_2}(a) and (b). Besides, the upper cutoff  $(r_{ucf}^{EN})$ is observed to shrink down with thermal populations in case of non-identical filters in Fig. \ref{EntVSthrmPop_nI0_5tauI2_5wK1_2}(b) which can be anticipated from Eq. \eqref{upper_lower_limit_thermal}. The entanglement is maximized for the input squeezing parameter $(r^{EN}_{max})$ when
	
	\begin{equation}\label{maximum_ent_thermal}
		\cosh(2r_{max}^{EN}) = \frac{\sqrt{A^2 B^2 \left( L_f^2-1\right)+A^4 K_f^2}}{\sqrt{\left(A^2 K_f^2+B^2 L_f^2\right) \left(A^2 \left(1-K_f^2\right)-B^2 L_f^2\right)}}
	\end{equation}

	While estimating two-mode squeezing, the maximally squeezed quadrature is realized for $\phi_I+\phi_S = \pi - \zeta$, where $\zeta = \arctan(\frac{BL_f}{AK_f})$. One can notice that the squeezing angle shifts with the change in the thermal population of the initial thermal light and the filter parameters. The thermal population can be dependent on the temperature and the frequency of interest. The angle of the filtered quadratue remains unchanged to the initial squeezing angle (i.e. $\zeta = 0$) only when the thermal population of both the reservoirs matches ($n_1 = n_2$) or two filters are identical ($L_f= 0$). The angle of the maximally squeezed quadrature is plotted in Fig. \ref{EntVSthrmPop_nI0_5tauI2_5wK1_2}(c) where the squeezed quadrature exhibits a clear dependence on thermal population. 
	
	The squeezing of the filtered arbitrary composite quadrature $ S_q (	X^{KL(\phi_I\phi_S)}_{IS} )$ given in Eq. \eqref{composite_quadrature} shows to have boundaries of the lower $(r^{SQ}_{lcf})$ and upper $(r^{SQ}_{ucf})$ cutoff limits within which the squeezing goes below SQL, where
	
	\begin{subequations}
		\begin{align}
			\tanh (r^{SQ}_{lcf}) = \frac{\sqrt{A^2 K_f^2+B^2 L_f^2}-\sqrt{A^2 K_f^2-A^2+B^2 L_f^2+1}}{A+1}  \\
			\tanh (r^{SQ}_{ucf}) = \frac{\sqrt{A^2 K_f^2+B^2 L_f^2}+\sqrt{A^2 K_f^2-A^2+B^2 L_f^2+1}}{A+1},
		\end{align}
	\end{subequations}
	
	The squeezing hits the standard quantum limit. 
	The squeezing reaches to its maximum for $r^{SQ}_{max}$, where
	
	\begin{equation}
		\tanh (2r^{SQ}_{max}) = \frac{1}{A}\sqrt{A^2 K_f^2 + B^2 L_f^2}
	\end{equation}

	In case of weighted quadratures $(X^{KL(\phi_I\phi_S)}_{IS(\mu_I\mu_S)})$ defined in Eq. \eqref{weighted_quadrature}, accepting the optimized weight  (see Appendix \ref{optimization_squeezing}), the maximally squeezed quadrature becomes
	
	\begin{multline}
		S_q(X^{KL}_{IS})_{optimized} =A \cosh (2 r)\\
		-\sqrt{\sinh ^2(2 r) \left(A^2 K^2+B^2 L^2\right)+B^2}
	\end{multline}

	The lower and upper cutoff boundaries of the input squeezing are obtained to be $(r^{EN}_{lcf})$ and $(r^{EN}_{ucf})$ determined in Eq. \eqref{upper_lower_limit_thermal}, for the maximally optimized squeezed quadratures within which the squeezing goes beyond SQL. The lower and upper cutoff borders are exhibited in Fig \ref{EntVSthrmPop_nI0_5tauI2_5wK1_2}(b,II), where we see the region of squeezing follows exactly same to the region of entanglement in Fig \ref{EntVSthrmPop_nI0_5tauI2_5wK1_2}(b,I), and therefore, reduces with the increment of thermal population. Also, one can obtain maximum squeezing for the value of input squeezing $r^{EN}_{max}$ given in Eq. \eqref{maximum_ent_thermal}.

	\subsection{Purity of the State}
	
	We estimate the purity of the two-mode squeezed thermal light in Fig. \ref{purity_thermal} and notice that the purity of the state goes down with an increasing thermal population. However, the purity remains unchanged with increasing input squeezing for identical filters. Anticipating that, furthermore, it is also observed to be destroyed with increasing input squeezing when the filters are non-identical. 
	
	
	\subsection{Quantum Non-Locality}
	
	We determine the region of quantum non-locality for the TMS thermal state in Fig. \ref{Bell_Thermal}. Following entanglement limits, in the case of identical filters, Fig. \ref{Bell_Thermal}(a) only exhibits a lower boundary of non-locality. Anticipating that, however, unlike Fig. \ref{EntVSthrmPop_nI0_5tauI2_5wK1_2}(a), the boundary of nonlocality in Fig. \ref{Bell_Thermal}(a) seems to reach a plato of the input thermal population, while increasing initial degree of squeezing ($r$). Besides, following entanglement limits in Fig. \ref{EntVSthrmPop_nI0_5tauI2_5wK1_2}(b), the non-locality seems to show an upper limit for non-identical filters in Fig. \ref{Bell_Thermal}(b), leaving a tiny region for non-local measurements.
	Moreover, the observation concludes that the region of quantum non-locality ensures entanglement, but the entanglement does not ensure non-locality, which can be justified from the mixness of the state.
	Rapid destruction of non-locality due to the influence of thermal bath was witnessed before in \cite{TMSV_nonlocal_Bell_Wigner}, however, the impact of filter has not been discussed, which is primarily covered here.

	\section{CONCLUSION}
	
	We studied the impacts of filters on TMS states and realized that the entanglement is maximized when the filters are identical. For non-identical filters, the entanglement goes down with the further increment of input squeezing, making a bell-like shape. We determined the condition where the entanglement is maximized and the cutoff limits, and realized that the region of entanglement does not change with the change of detection efficiency, even after applying filter. 
	While investigating two-mode squeezed thermal light, the entanglement is observed to be robust against the thermal population. 
	As a complementary estimation of logarithmic negativity, we determined the TMS of optimized quadratures, which can be experimentally measured through homo/heterodyne detection. 
	The SQL limited boundaries and the maximal squeezing follows the entanglement profile, as it is a direct measurement of corresponding logarithmic negativity \cite{Vitali_Zippilli_NJP}. 	
	Together with the non-identicality of filters, the temperature of thermal light manipulates the angle of maximally squeezed quadrature. 
	The non-identicality of the filter ensures the state to be mixed.
	Following the entanglement profile, we also realize that the filter causes limitations in the non-local measurement of the filtered output. 	
	However, unlike entanglement, the region of non-locality is observed to reduce with the reduction of detection efficiency. 
	More importantly, since the states are mixed, the violation of Bell inequality occurs if and only if the state is entangled. Therefore, the entanglement becomes a necessary criterion for non-locality, but not sufficient.	
	The filter affected destruction of entanglement and non-locality at a higher degree of input squeezing, ensuring to
	conclude that while performing quantum optical experiments, the generation of a two-mode highly squeezed state may not be beneficial for the generation of entanglement and non-local measurement for the experiments related to quantum information, communication, or metrology.

	\begin{acknowledgments}
		SA would like to thank Philippe Djorwe for his fruitful suggestions. The work has been supported by the European Union, MSCA GA no 101065991 (SingletSQL).
		
	\end{acknowledgments}

	\appendix
	
	\section{Optimization of Maximally Squeezd Quadrature}\label{optimization_squeezing}
	
	We estimated the parameter space here where the fluctuation of hybrid quadrature can reach beyond SQL. 
	The fluctuation of a maximally squeezed weighted quadrature is 
	
	\begin{multline}
		S_q(	X^{KL}_{IS(\mu_I\mu_S)}) =\frac{1}{\mu_I^2+\mu_S^2} \bigg[ \mu_I^2  D_{I}  +  \mu_S^2 D_{S} \\
		- 2\mu_I\mu_S \sqrt{C_{11}^2  +   C_{21}^2  }
		\bigg]
	\end{multline}
	
	The optimized weight ratio for which the squeezing is maximum
	
	\begin{equation}
		\frac{\mu_I}{\mu_S} =\frac{D_S - D_I + \sqrt{4 C_{11}^2 + 4 C_{12}^2 + D_I^2 - 2 D_I D_S + D_S^2} }{2 \sqrt{C_{11}^2 + C_{12}^2}}
	\end{equation}
	
	Accepting the optimized weight ratio, the maximal TMS quadrature gives
	
	\begin{multline}
		S_q(	X^{KL}_{IS(\mu_I\mu_S)})|_{optimized} = \\
		\frac{1}{2} (D_I + D_S - \sqrt{4 C_{11}^2 + 4 C_{12}^2 + (D_I - D_S)^2}  )
	\end{multline}

	\subsection{Maximally Squeezd TMSV Quadrature}

	The minimized fluctuation of the hybrid quadrature $S_q(X^{KL(\phi_I\phi_S)}_{IS(\mu_I\mu_S)})$ at $\phi_I+\phi_S = \pi$ can further be optimized for the ratio of weights
	
	\begin{multline}\label{weight_ratio}
		\frac{\mu_I}{\mu_S}  =  \\
		\frac{ \sqrt{4 \eta _I \eta _S {K_f}^2 +(\eta _I-\eta _S)^2 \tanh ^2(r)}+(\eta _S-\eta _I) \tanh (r) }{2 {K_f} \sqrt{\eta _I \eta _S}}.
	\end{multline}
	
\begin{figure}[t!]
		\includegraphics[width= 1 \linewidth]{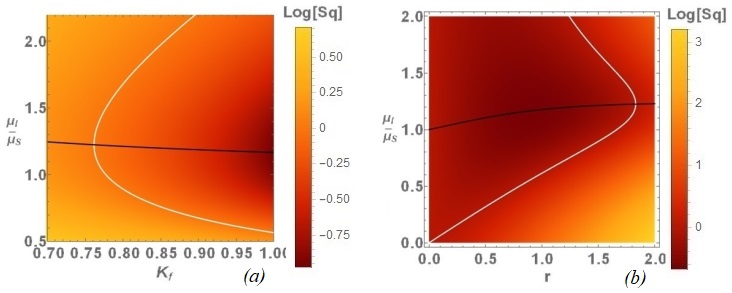}
		\caption{ Squeezing of two parties filtered output for the variation of (a) the weight ratio $\mu_I/\mu_S$ vs $K_f$ when $r=1$ and (b)  $\mu_I/\mu_S$ vs $r$ when $K_f=0.95$.
			The fixed optical quantum efficiencies $ \eta_I = 0.6, \eta_S = 0.9$. 	
		}\label{ENSQTMSV_test}
	\end{figure}

	Fig. \ref{ENSQTMSV_test}(a) shows how the squeezing of maximum squeezed quadrature varies with the change of $K_f$ and the ratio $\mu_I/\mu_S$. Based on that, we optimized maximal squeezing (which is plotted by black lines) which remains to be $\mu_I/\mu_S >1$ for $\eta_S >\eta_I$. The weight ratio reduces approaching to be unit valued when $K_f$ increases to 1. We also determined how the squeezing varies with the change of $r$ and the ratio $\mu_I/\mu_S$ in Fig. \ref{ENSQTMSV_test}(b), and realized an opposite phenomenon, that the weight ratio for the maximally optimized squeezing increases from its unit value when $r$ increases. We also estimated the borderline where the squeezing goes beyond the SQL for both the cases 
	.

	\subsection{Maximally Squeezd TMS Thermal Quadrature}

	In case of weighted quadratures $(X^{KL(\phi_I\phi_S)}_{IS(\mu_I\mu_S)})$ defined in Eq. \eqref{weighted_quadrature}, 	
	the squeezing $S_q(X^{KL(\phi_I\phi_S)}_{IS(\mu_I\mu_S)})$ is maximized for the weight ratio

	\begin{equation}\label{thermal__weight}
		\frac{\mu_I}{\mu_S} = \frac{ \sqrt{ \sinh ^2(2 r) \left(A^2 K_f^2+B^2 L_f^2\right)+ B^2}- B }{ \sinh (2 r) \sqrt{A^2 K_f^2+B^2 L_f^2}}
	\end{equation}

\begin{figure}
		\includegraphics[width= 1 \linewidth]{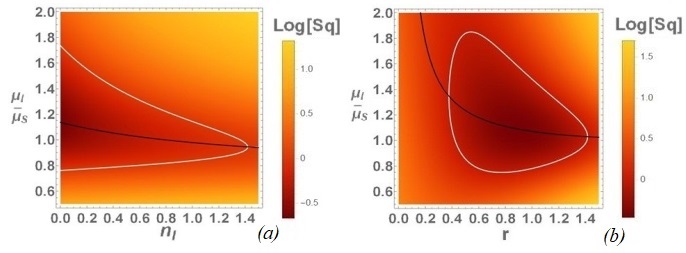}
		\caption{  Squeezing of two parties filtered output for the variation of (a) the weight ratio $\mu_I/\mu_S$ vs $n_I$ when $r=1$ and (b)  $\mu_I/\mu_S$ vs $r$ when $n_I=0.3$.
		For both the cases $n_S= 0.8$,  $K_f = 0.95$ and $L_f = 0.095 $ are the fixed parametrs
			and the optical quantum efficiencies $(\eta_{I,S})$ are considered to be 1. }\label{ThermalPop_K0_95L0_095}
	\end{figure}

	We visualize how the combined quadrature variance changes with the variation of the thermal population in Fig \ref{ThermalPop_K0_95L0_095}(b-I) and input squeezing factor in Fig \ref{ThermalPop_K0_95L0_095}(b-II). The ratio reduces with the increment of the thermal population ($n_I$) and the weights become equal when the population matches $(n_I =n_S)$. Furthermore, we also realize the weights approach towards equality with the increment of input squeezing, which can also be hinted from the Eq. \eqref{thermal__weight}. In both cases, we also determined the boundaries beyond which the squeezing below SQL disappears.

	\nocite{*}

	\bibliography{apssamp}
	\bibliographystyle{apsrev4-2}

\end{document}